\documentclass[12pt]{article}
\usepackage{amsfonts}
\begin{document}
\begin{center}

\textbf{CARDY-VERLINDE ENTROPY FORMULA IN THE PRESENCE OF A GENERAL COSMOLOGICAL STATE EQUATION}\\
\vspace{1cm}
I. Brevik\footnote{E-mail address: iver.h.brevik@mtf.ntnu.no}\\
\vspace{1cm}
Division of Applied Mechanics,\\Norwegian University of Science and Technology,\\
N-7491 Trondheim, Norway\\
\vspace{1cm}
April 2002
\vspace{1cm}
\end{center}
\begin{abstract}

As recently found by Youm [hep-th/0201268], the entropy of the universe will no longer be expressible in the conventional Cardy-Verlinde form if one relaxes the radiation dominance state equation and instead assumes a more general equation of the form $p=(\gamma-1)\rho$, with $\gamma$ a constant. We show that Youm's generalized entropy formula remains valid when the cosmic fluid is no longer ideal, but is allowed to possess a constant bulk viscosity $\zeta$. We supply our analysis with some numerical estimates, thus calculating the scale factor $a(t)$ for a $k=0$ universe, and also calculate via a perturbative expansion in $\zeta$ the magnitude of the viscosity-induced correction to the scale factor if the universe is radiation dominated.

\end{abstract}
\vspace{1cm}
\newpage

\section{Introduction}

The suggestion of Verlinde \cite{verlinde00} that there may exist a holographic bound on the subextensive entropy associated with the Casimir energy, makes it natural to ask: is this formal merging between the holographic principle, the Cardy entropy formula from conformal field theory \cite{cardy86}, and the Friedmann equations from cosmology, merely a formal coincidence or does it reflect a deep physical property? As one might expect, the Verlinde proposal has inspired a considerable number of investigations, wherein various facets of the Cardy-Verlinde entropy formula, in particular, have been elucidated. An extensive list of references to the recent literature can be found in the paper of Youm \cite{youm02}. Let us only mention here the large interest that  has arisen in connection with the entropy and the energy as following from quantum and thermal fluctuations in conformal field theories \cite{kutasov01}.

Verlinde assumed a spatially closed and radiation dominated universe. It becomes natural to inquire to what degree these basic assumptions can be generalized. For instance, Wang {\it et al.} \cite{wang01} considered universes having a cosmological constant, and Nojiri {\it et al.} \cite{nojiri01} considered quantum bounds for the Cardy-Verlinde formula. Recently, we examined the generalization of the Cardy-Verlinde formula allowing for a constant {\it bulk viscosity} $\zeta$ in the radiation dominated universe \cite{brevik02}. The conclusion of our analysis was that such a generalization is mathematically possible, although one should bear in mind that there occurs a possible interpretional difficulty: it is not evident from a physical point of view that the interpretiation of the subextensive part of the energy should be interpreted as a {\it Casimir energy}.

The mentioned paper of Youm \cite{youm02} is interesting, since it shows that the entropy of the universe can no longer be expressed in the conventional Cardy-Verlinde form if one relaxes the radiation dominance state equation and instead assumes a more general equation
\begin{equation}
p=(\gamma-1)\rho,
\end{equation}
\label{1}
with $\gamma$ a constant. A more general form of the entropy formula results. And this brings us to the main theme of the present note, namely to explore to which extent the presence of a bulk viscosity in the cosmic fluid, together with the state equation (1), influences the entropy formula. In fact, it will turn out that the modified entropy formula found by Youm still persists, even when a constant bulk viscosity is  allowed for. The central formula is Eq.~(15) below.

We also supply the general analysis with some numerical estimates on viscous cosmology. Thus Section 3.1 calculates the scale factor $a(t)$ for a viscous universe, assuming for simplicity that $k=0$, if the viscosity is taken to set in suddenly at some instant $t=t_{\rm in}$, which we choose to lie in the plasma era. Section 3.2 exploits the smallness of $\zeta$ to calculate the viscosity-induced correction to $a(t)$ via a perturbative expansion in $\zeta$ to the lowest order. The correction becomes very small, if $\zeta$ is calculated in the plasma era from conventional kinetic theory. However, at earlier instants in the history of the universe the value of $\zeta$ might have been  much higher, thus strengthening the effect.

\section{Inclusion of viscosity}

We take the number $n$ of spatial dimensions to be equal to 3, assume a FRW metric with k=+1,
\begin{equation}
ds^2=-dt^2+a^2(t)\left( \frac{dr^2}{1-r^2}+r^2d\Omega^2\right),
\end{equation}
\label{2}
set the cosmological constant $\Lambda$ equal to zero, and assume that the cosmic fluid possesses a constant bulk viscosity $\zeta$. The energy-momentum tensor can then be written as
\begin{equation}
T_{\mu\nu}=\rho U_\mu U_\nu+(p-\zeta \theta)h_{\mu\nu},
\end{equation}
\label{3}
where in comoving coordinates $U^\mu=(1,0),~~\theta \equiv {U^\mu}_{;\mu}= 3\dot{a}/a$,  $h_{\mu\nu}=g_{\mu\nu}+U_{\mu} U_\nu$ being the projection tensor. The effective pressure is $\tilde{p}=p-\zeta \theta$. From Einstein's equations $R_{\mu\nu}-\frac{1}{2}Rg_{\mu\nu}=8\pi G T_{\mu\nu}$ we derive the first Friedmann equation (the "initial value equation")
\begin{equation}
H^2=\frac{8\pi G}{3}\rho-\frac{1}{a^2}
\end{equation}
\label{4}
(with $H=\dot{a}/a$), which is seen to be independent of the viscosity. The second Friedmann equation (the "dynamic equation"), when combined with Eq.~(4), yields
\begin{equation}
\dot{H}=-4\pi G(\rho+\tilde{p})+\frac{1}{a^2},
\end{equation}
\label{5}
in which the effect from viscosity is explicit. The conservation equation for energy, ${T^{0\nu}}_{;\nu}=0$, yields
\begin{equation}
\dot{\rho}+(\rho+p)\theta=\zeta \theta^2.
\end{equation}
\label{6}
A direct calculation using the Friedmann equations leads to the equation
\begin{equation}
\frac{d}{da}\left( \rho a^4\right) =(\rho-3\tilde{p})a^3.
\end{equation}
\label{7}
So far, the equation of state for the cosmic fluid has not been used. Let us hereafter assume the specific form (1). Then, using Eq.~(7) we derive
\begin{equation}
\frac{d}{dt}\left( \rho a^{3\gamma}\right)=\zeta\,\theta^2a^{3\gamma}.
\end{equation}
\label{8}
This equation is useful for our purpose. We wish to combine it with the conservation law for entropy: if $n$ is the particle number density and $\sigma$ the entropy per particle, the four-divergence of the entropy current vector, $S^\mu=n\sigma U^\mu$, is
\begin{equation}
{S^\mu}_{;\mu}=\frac{\zeta}{T}\theta^2
\end{equation}
\label{9}
which, in view of the equation $(nU^\mu)_{;\mu}=0$, yields
\begin{equation}
n\dot{\sigma}=\frac{\zeta}{T}\theta^2
\end{equation}
\label{10}
in the comoving coordinate system.

We can now carry out the same kind of reasoning as in Ref.~\cite{brevik02}: Since $\zeta$ is small, we can use for $a=a(t)$ the same expression as for a nonviscous closed universe, namely  $a(t)=a_* \sin \eta$, where $\eta$ is the conformal time, the constant $a_*$ is defined as
\begin{equation}
a_* =\sqrt{{(8\pi G/3)\rho_{\rm in}a_{\rm in}^4}}, 
\end{equation}
\label{11}
and the subscript "in" designates the initial instant of the onset of viscosity. (Strictly speaking, the approximate expression (11) assumes that the universe is radiation dominated.) 
 Imagine that Eqs~(8) and (10) are integrated with respect to time. Since $\zeta^{-1}\rho a^{3\gamma}$ and $\zeta^{-1}n\sigma $ can be drawn as functions of $t$, it follows that $\rho a^{3\gamma}$ can be considered as a function of $n\sigma$. Then, since the total energy is $E \sim \rho a^3$ and the total entropy is $S \sim n\sigma a^3$, it follows that $E a^{3(\gamma-1)}$ is independent of the volume $V$ and is a function of $S$ only. This generalizes the pure entropy dependence of the product $Ea$, found by Verlinde \cite{verlinde00} in the case of a nonviscous radiation dominated universe. And it is noteworthy that the derived property of $Ea^{3(\gamma-1)}$ formally agrees exactly with the property found by Youm \cite{youm02} when $\zeta=0$.

Let us carry out the analysis a bit further, and write the total energy $E$ as a sum of an extensive part $E_E$ and a subextensive part $E_C$:
\begin{equation}
E(S,V)=E_E(S,V)+\frac{1}{2}E_C(S,V).
\end{equation}
\label{12}
Under a scale transformation $S \rightarrow \lambda S$ and $ V \rightarrow \lambda V $ with constant $\lambda$, $E_E$ scales linearly with $\lambda$. But the term $E_C$ scales with a power of $\lambda$ that is less than one: as $E_C$ is the volume integral over a local energy density expressed in the metric and its derivatives, each of which scales as $\lambda^{-1/3}$, and as the derivatives occur in pairs, the power in $\lambda$ has to be 1-2/3= 1/3. Thus we have
\begin{equation}
E_E(\lambda S, \lambda V)=\lambda E_E(S,V), \quad E_C(\lambda S,\lambda V)=\lambda^{1/3}\,E_C(S,V),
\end{equation}
\label{13}
which implies
\begin{equation}
E_E=\frac{C_1}{4\pi a^{3(\gamma-1)}}\,S^\gamma,\quad E_C=\frac{C_2}{2\pi a^{3(\gamma-1)}}\,S^{\gamma-2/3},
\end{equation}
\label{14}
where $C_1, C_2$ are constants. In CFTs, their product is known: $\sqrt{C_1C_2}=n=3$ \cite{verlinde00}; this being a consequence of the AdS-CFT correspondence. Using Eqs.~(12) and (14) we obtain
\begin{equation}
S=\left[ \frac{2\pi a^{3(\gamma-1)}}{\sqrt{C_1C_2}}\sqrt{E_C(2E-E_C)}\right]^{\frac{3}{3\gamma-1}}.
\end{equation}
\label{15}
This is the generalized Cardy-Verlinde formula, in agreement with Eq.~(20) in Youm's paper, reducing to the standard formula (with square root) in the case of a radiation dominated universe. In conclusion, we have extended the basis of Eq.~(15) so as to include the presence of a constant bulk viscosity in the cosmic fluid.

\section{Numerical estimates}

\subsection{Spatially flat universe}

It is of physical interest to supplement the above considerations with some simple numerical estimates, showing, in particular, the order of magnitude of the viscosity terms. Let us assume, as mentioned above,  that the effect of the bulk viscosity effectively sets in at some initial instant $t=t_{\rm in}$ and that $\zeta$ is thereafter constant for the times that we consider. For definiteness we choose
\begin{equation}
t_{\rm in}=1000~{\rm s}
\end{equation}
\label{16}
after the big bang. The universe is then in the plasma (or radiation) era; it consists of ionized H and He in equilibrium with radiation. The particle density is $n_{\rm in}=10^{19}~{\rm cm^{-3}}$, and the temperature is $T_{\rm in}=4\times 10^8$ K. The advantage of considering this relatively late stage of the universe's history is that the magnitude of $\zeta$ can be calculated using conventional kinetic theory. At $t=t_{\rm in}$ one finds
\begin{equation}
\zeta=7.0\times 10^{-3}~{\rm g\,cm^{-1}\, s^{-1}}
\end{equation}
\label{17}
(cf. \cite{brevik94} and further references therein).

For our estimate purposes it appears reasonable to assume that the influence from the spatial curvature is not very important. For simplicity let us put $k=0$. This implies that there exists the following simple differential equation for the scalar expansion \cite{brevik94}:
\begin{equation}
\dot{\theta}(t)+\frac{1}{2}\gamma \,\theta^2(t)-12\pi G \zeta \theta(t)=0,
\end{equation}
\label{18}
which can be solved to give the expression for the scale factor:
\begin{equation}
a(t)=a_{\rm in}\left[ 1+\frac{1}{2}\gamma\, \theta_{\rm in} t_c\left( e^{(t-t_{\rm in})/t_c}-1\right) \right]^{2/(3\gamma)},
\end{equation}
\label{19}
where
\begin{equation}
t_c=(12\pi G\zeta)^{-1}.
\end{equation}
\label{20}.
The corresponding expression when viscosity is absent, is
\begin{equation}
a(t,\zeta=0)=a_{\rm in}\left[ 1+\frac{1}{2}\gamma \theta_{\rm in}(t-t_{\rm in})\right]^{2/(3\gamma)}.
\end{equation}
\label{21}
The ratio between the expressions (19) and (21) reduces to unity in the limit when $(t-t_{\rm in})/t_c \ll 1$. This is we would expect. The influence from viscosity generally turns up only in the factor $t_c$, and the effect becomes strengthened when $t_c$ becomes smaller, {\it i.e.}, when $\zeta$ becomes larger. The numerical values given above for the instant $t_{\rm in}=1000$ s correspond to
\begin{equation}
\theta_{\rm in}=1.5\times 10^{-3}~{\rm s^{-1}},\quad t_c=5.1\times 10^{28}~{\rm s}.
\end{equation}
\label{22}

\subsection{ Perturbative expansion for a radiation dominated closed universe}

The smallness of $\zeta$ makes it natural, as an alternative to the approach of the previous subsection, to make a Stokes expansion in $\zeta$. For simplicity we now assume that $\gamma =4/3$, {\it i.e.}, that the universe is radiation dominated. We put $k=1.$ Let subscript zero refer to quantities in the nonviscous case. We write the first order expansions as
\begin{equation}
a=a_0(1+\zeta a_1), \quad \rho=\rho_0 (1+\zeta \rho_1),
\end{equation}
\label{23}
where the functions $a_1$ and $\rho_1$ are of zeroth order in $\zeta$; they are regarded as functions of $t$ or alternatively as functions of the conformal time $\eta$. Correspondingly, the scalar expansion $\theta =3\dot{a}/a$ is written as
\begin{equation}
\theta=\theta_0(1+\zeta \theta_1).
\end{equation}
\label{24}
The Friedmann equation (4), and Eq.~(8), now yield to the zeroth order
\begin{equation}
\theta_0^2=24\pi G\rho_0-\frac{9}{a_0^2},
\end{equation}
\label{25}
\begin{equation}
\rho_{\rm in}a_{\rm in}^4=\rho_0 a_0^4,
\end{equation}
\label{26}
and to the first order
\begin{equation}
\theta_0^2\,\theta_1=12\pi G\rho_0\rho_1+\frac{9a_1}{a_0^2},
\end{equation}
\label{27}
\begin{equation}
\rho_{\rm in}a_{\rm in}^4(\dot{\rho}_1+4\dot{a}_1)=\theta_0^2\, a_0^4.
\end{equation}
\label{28}
The solutions of Eqs.~(25) and (26) are
\[ a_0=a_* \sin \eta, \quad \rho_0=\frac{3}{8\pi G}\,\frac{1}{a_*^2}\,\frac{1}{\sin^4 \eta}, \]
\begin{equation}
\theta_0=\frac{3\cos \eta}{a_*\sin^2\eta},
\end{equation}
\label{29}
where $a_*$ is defined by Eq.~(11). From Eq.~(28) we now have
\begin{equation}
\frac{d\rho_1}{d\eta}+\frac{4da_1}{d\eta}=24\pi Ga_*\sin \eta \cos^2\eta,
\end{equation}
\label{30}
which can be integrated from $\eta=\eta_{\rm in}$ onwards:
\begin{equation}\rho_1+4a_1=8\pi Ga_*(\cos^3\eta_{\rm in}-\cos^3\eta).
\end{equation}
\label{31}
We have here assumed that $a_1=\rho_1=0$ at the initial instant $t=t_{\rm in}$. With
\begin{equation}
\theta_1=\frac{da_1}{d\eta}\tan \eta
\end{equation}
\label{32}
(cf. Eq.~(23)), we obtain from Eq.~(27) a first order differential equation for the scale factor correction:
\begin{equation}
\sin 2\eta\,\frac{da_1}{d\eta}+2(1+\cos^2\eta)a_1+8\pi Ga_*\cos^3 \eta=8\pi Ga_*\cos^3\eta_{\rm in}.
\end{equation}
\label{33}
This equation admits the integrating factor \cite{korn68}
\begin{equation}
\mu(\eta)=\exp \left[ \int \frac{2(1+\cos^2\eta)}{\sin 2\eta}\, d\eta\right]=\sqrt{2}\,\frac{\sin^2\eta}{\cos\eta},
\end{equation}
\label{34}
whereby we find after some calculation the following solution, again observing the initial conditions at $t=t_{\rm in}$:
\begin{equation}
a_1(\eta)=\frac{4\pi Ga_*}{\sin^2\eta}\left[ \cos^3\eta_{\rm in}+\left( \frac{1}{4}\cos 2\eta-\cos^2\eta_{\rm in}-\frac{1}{4}\cos 2\eta_{\rm in}\right)\cos \eta \right].
\end{equation}
\label{35}
Once the scale factor correction $a_1$ is known, the corresponding density correction $\rho_1$ follows from Eq.~(31).

From Eq.~(35) it is apparent that the relative correction $\zeta a_1$ to the scale factor is of order $4\pi G\zeta a_*$ or, in dimensional units, $4\pi G\zeta a_*/c^3$. We here note that $4\pi G\zeta/c^2=1/(3t_c)$, thus about $6.5 \times 10^{-30}~{\rm s}^{-1}$ according to Eq.~(20), which in dimensional form reads $t_c^{-1}=12\pi G \zeta/c^2$. The quantity $a_*$, according to Eq.~(29), is the maximum value of the scale factor in a nonviscous $k=1$ theory. Let us put $a_*$ equal to the commonly accepted value of the radius of the universe, {\it i.e.,} $a_*=10^{28}$ cm. Then we obtain 
  $4\pi G\zeta a_*/c^3=2\times 10^{-12}$. The relative correction to the scale factor is thus in this case very small. Physically, this is due to the fact that we are considering an instant relatively late in the history of the universe. If the bulk viscosity $\zeta$ were higher at earlier times, or, if there were an {\it anisotropic} stage present in the early universe at which the enormously higher {\it shear} viscosity would come into play \cite{brevik94}, then the effect would be significantly enhanced.

\vspace{1cm}
{\bf Acknowledgment}

\bigskip
I thank professor Sergei Odintsov for valuable correspondence.


\newpage

\begin{thebibliography}{99}

\bibitem{verlinde00}
E. Verlinde, hep-th/0008140.

\bibitem{cardy86}
J. L. Cardy, Nucl. Phys. B {\bf 270}, 186 (1986); {\it ibid.} B {\bf 275}, 200 (1986).

\bibitem{youm02}
D. Youm, hep-th/0201268.

\bibitem{kutasov01}
D. Kutasov and F. Larsen, JHEP {\bf 0101}, 001 (2001) [hep-th/0009244]; D. Klemm, A. C. Petkou, and G. Siopsis, Nucl. Phys. B {\bf 601}, 380 (2001); S. Nojiri and S. D. Odintsov, Int. J. Mod. Phys. A {\bf 16}, 3273 (2001); Class. Quant. Grav. {\bf 18}, 5227 (2001); S. Nojiri, S. D. Odintsov, and S. Ogushi, Int. J. Mod. Phys. A {\bf 16}, 5085 (2001);
J. E. Lidsey, S. Nojiri, and S. D. Odintsov, hep-th/0202198; A. Momen and T. Sarkar, hep-th/0203244;
I. Brevik, K. A. Milton, and S. D. Odintsov, hep-th/0202048 v3.

\bibitem{wang01}
B. Wang, E. Abdalla, and R.-K. Su, Phys. Lett. B {\bf 503}, 394 (2001).

\bibitem{nojiri01}
S. Nojiri, O. Obregon, S. D. Odintsov, H. Quevedo, and M. P. Ryan, Mod. Phys. Lett. A {\bf 16}, 1181 (2001).

\bibitem{brevik02}
I. Brevik and S. D. Odintsov, Phys. Rev. D {\bf 65}, 067302 (2002).

\bibitem{brevik94}
I. Brevik and L. T. Heen, Astrophys. Space Sci. {\bf 219}, 99 (1994).

\bibitem{korn68}
See, for instance, G. A. Korn and T. M. Korn, {\it Mathematical Handbook for Scientists and Engineers}, 2nd ed. (McGraw-Hill, New York, 1968), p. 250.



\end{thebibliography}
\end{document}